\documentclass[11pt]{article}

\usepackage{times}
\usepackage{amsfonts}
\usepackage{amssymb}
\usepackage{amsmath}
\usepackage{fullpage}

\newcommand{\ket}[1]{|\hspace{0.5pt}#1\hspace{0.5pt}\rangle}

\newcommand{\bra}[1]{\langle\hspace{0.5pt}#1\hspace{0.5pt}|}
\newcommand{\bracket}[2]{\langle\hspace{0.5pt}#1\hspace{0.5pt}|\hspace{0.5pt}
	#2\hspace{0.5pt}\rangle}
\newcommand{\opbracket}[3]{\langle\hspace{0.5pt}#1\hspace{0.5pt}|\hspace{0.5pt}
	#2\hspace{0.5pt}|\hspace{0.5pt}#3\hspace{0.5pt}\rangle}

\newcommand{\op}[1]{\operatorname{#1}}
\def \qed {\hfill \rule{2mm}{2mm}\vspace{3mm}}

\newenvironment{mylist}[1]{\begin{list}{}{
	\setlength{\leftmargin}{#1}
	\setlength{\rightmargin}{0mm}
	\setlength{\labelsep}{2mm}
	\setlength{\labelwidth}{8mm}
	\setlength{\itemsep}{0mm}}}
	{\end{list}}
\newtheorem{theorem}{Theorem}
\newtheorem{lemma}[theorem]{Lemma}
\newtheorem{prop}[theorem]{Proposition}

\begin{document}

\title{\Large\bf Notes on super-operator norms induced by Schatten norms}

\author{John Watrous\\[2mm]
  Institute for Quantum Information Science\\
  and Department of Computer Science\\
  University of Calgary\\
  Calgary, Alberta, Canada
}

\date{November 10, 2004}

\maketitle

\begin{abstract}
Let $\Phi$ be a super-operator, i.e., a linear mapping of the form
$\Phi:\mathrm{L}(\mathcal{F})\rightarrow\mathrm{L}(\mathcal{G})$ for
finite dimensional Hilbert spaces $\mathcal{F}$ and $\mathcal{G}$.
This paper considers basic properties of the super-operator norms defined by
$\|\Phi\|_{q\rightarrow p}= \sup\{\|\Phi(X)\|_p/\|X\|_q\,:\,X\not=0\}$,
induced by Schatten norms for $1\leq p,q\leq\infty$.
These super-operator norms arise in various contexts in the study of quantum
information.

In this paper it is proved that if $\Phi$ is completely positive, the value of
the supremum in the definition of $\|\Phi\|_{q\rightarrow p}$ is achieved by a
positive semidefinite operator $X$, answering a question recently posed by
King and Ruskai~\cite{KingR04}.
However, for any choice of $p\in [1,\infty]$, there exists a super-operator
$\Phi$ that is the {\em difference} of two completely positive,
trace-preserving super-operators such that all Hermitian $X$ fail to achieve
the supremum in the definition of $\|\Phi\|_{1\rightarrow p}$.

Also considered are the properties of the above norms for super-operators
tensored with the identity super-operator.
In particular, it is proved that for all $p\geq 2$, $q\leq 2$, and arbitrary
$\Phi$, the norm $\|\Phi \|_{q\rightarrow p}$ is stable under tensoring $\Phi$
with the identity super-operator, meaning that
$\|\Phi \|_{q\rightarrow p} = \|\Phi \otimes I\|_{q\rightarrow p}$.
For $1\leq p < 2$, the norm $\|\Phi\|_{1\rightarrow p}$ may fail to be stable
with respect to tensoring $\Phi$ with the identity super-operator as just
described, but $\|\Phi\otimes I\|_{1\rightarrow p}$ is stable in this sense
for $I$ the identity super-operator on $\mathrm{L}(\mathcal{H})$
for $\op{dim}(\mathcal{H}) = \op{dim}(\mathcal{F})$.
This generalizes and simplifies a proof due to Kitaev \cite{Kitaev97} that
established this fact for the case $p=1$.
\end{abstract}


\section{Introduction}

Super-operators, which are linear mappings from square matrices to
square matrices, play an important role in quantum information theory.
For instance, discrete physical changes in quantum systems, such as
computations and noise, are represented by completely positive trace
preserving super-operators.
Another example is the partial transpose, which is a non-positive
super-operator that is very important in the theory of entanglement.
This paper concerns basic properties of a class of norms defined on
super-operators.

Let $\mathcal{F}$ and $\mathcal{G}$ be finite-dimensional complex vector
spaces, and let $\mathrm{L}(\mathcal{F},\mathcal{G})$ denote the vector
space consisting of all linear mappings from $\mathcal{F}$ to $\mathcal{G}$.
We will write $\mathrm{L}(\mathcal{F})$ as shorthand for
$\mathrm{L}(\mathcal{F},\mathcal{F})$.
The space $\mathrm{L}(\mathcal{F},\mathcal{G})$ is an inner product space
with respect to the inner product defined by
$\displaystyle\langle X, Y\rangle = \op{tr}X^{\ast} Y$.

For any real number $p\geq 1$, the Schatten $p$-norm is a norm on
$\mathrm{L}(\mathcal{F})$ defined by
\[
\|X\|_p = \left(\op{tr}|X|^p\right)^{\frac{1}{p}}
\]
for every $X\in\mathrm{L}(\mathcal{F})$, 
where $|X| =\sqrt{\displaystyle X^{\ast}X}$.
Equivalently, $\|X\|_p$ is the usual $p$-norm of the vector of singular
values of $X$.
Also let
\[
\|X\|_{\infty} = \lim_{p\rightarrow \infty}\|X\|_p,
\]
which is equivalent to the usual operator norm of $X$.

The space $\mathrm{T}(\mathcal{F},\mathcal{G})$ consists of all
super-operators $\Phi$ of the form
$\Phi: \mathrm{L}(\mathcal{F}) \rightarrow \mathrm{L}(\mathcal{G})$.
For $1\leq p,q\leq \infty$, define the norm $\|\cdot\|_{q\rightarrow p}$ on
$\mathrm{T}(\mathcal{F},\mathcal{G})$ as
\[
\|\Phi\|_{q\rightarrow p} = \sup_{X\not=0}
\frac{\|\Phi(X)\|_p}{\|X\|_q},
\]
and let $\|\Phi\|_{q\rightarrow p}^H$ be defined similarly, except only
taking the supremum over Hermitian operators:
\[
\|\Phi\|_{q\rightarrow p}^H = \sup_{X=X^{\ast}\not=0}
\frac{\|\Phi(X)\|_p}{\|X\|_q}.
\]
Obviously $\|\Phi\|_{q\rightarrow p}^H \leq \|\Phi\|_{q\rightarrow p}$,
and strict inequality may occur for some choices of $\Phi$, $p$, and $q$.
The case where $q = 1$ is of particular importance in quantum information
theory, and we simply write $\|\Phi\|_p$ to denote $\|\Phi\|_{1\rightarrow p}$.

The above norms, and variants of these norms, arise in different contexts in
quantum information theory.
For instance, given two completely positive, trace-preserving super-operators
$\Phi_0$ and $\Phi_1$, the quantity\linebreak
$\|\Phi_0 - \Phi_1\|_1^H$ represents 
one way of measuring the distance between $\Phi_0$ and $\Phi_1$.
By a simple convexity argument,
\[
\|\Phi_0 - \Phi_1\|_1^H = \max\{\|\Phi_0(\ket{\psi}\bra{\psi}) - 
\Phi_1(\ket{\psi}\bra{\psi})\|_1\,:\,\|\ket{\psi}\| = 1\},
\]
and so this quantity characterizes the maximum probability of distinguishing
the outputs of $\Phi_0$ and $\Phi_1$ over all pure state inputs.
Kitaev \cite{Kitaev97} observed that this norm possesses some undesirable
properties when used to characterize distance in this way.
In particular, this norm is not stable with respect to taking tensor
products with identity super-operators, i.e., there exist choices for
completely positive trace-preserving $\Phi_0$ and $\Phi_1$ for which 
\[
\|\Phi_0 - \Phi_1\|_1^H < \|\Phi_0\otimes I - \Phi_1\otimes I\|_1^H,
\]
where $I$ denotes the identity super-operator over some ``ancilla'' space.
Operationally, this implies that distinguishing super-operators $\Phi_0$
and $\Phi_1$ may become easier if one applies $\Phi_0$ or $\Phi_1$ to one
part of an initially entangled bipartite system.
To remedy this situation Kitaev defined a stabilized version of the
1-norm as follows:
\[
\|\Phi\|_{\Diamond} = \|\Phi\otimes I\|_1.
\]
Here $I$ denotes the identity super-operator on a space of dimension
equal to that of the input space of $\Phi$.
Tensoring $\Phi$ with the identity operator on a larger space cannot increase
its 1-norm.
The norm $\|\cdot\|_{\Diamond}$ has some remarkable properties, and has
found use in various contexts such as in the study of quantum circuits,
error correction, and quantum interactive proof systems.

Amosov, Holevo, and Werner \cite{AmosovH+00} considered the quantity
$\|\Phi\|_p^H$ (and other related quantities) of a given completely positive
trace-preserving super-operator $\Phi$ as a measure of its highest output
purity, and conjectured that this quantity is multiplicative for completely
positive $\Phi$:
\begin{equation}
\label{eq:multiplicativity-conjecture}
\|\Phi\otimes\Psi\|_p^H \stackrel{?}{=} \| \Phi\|_p^H \|\Psi\|_p^H.
\end{equation}
This conjecture was later refuted by Werner and Holevo \cite{WernerH02}
for $p\geq 4.79\cdots$, but its status for smaller $p$ is still unknown.
The truth of the conjecture (\ref{eq:multiplicativity-conjecture})
for $p\in [1,1 + \varepsilon)$ for positive $\varepsilon$ would have
major consequences in quantum information theory.
In particular, it would imply the strong superadditivity of the entanglement
of formation \cite{AudenaertB04}, which is equivalent
to three other central additivity conjectures in quantum information
theory \cite{Shor04}.

Much of the recent work concerning the above conjecture has focused on
proving the conjecture for special classes of completely positive
super-operators, such as the case where one of the super-operators is
a depolarizing channel \cite{AmosovH03}, a unital qubit channel \cite{King02},
or an entanglement-breaking channel \cite{King03}.
King and Ruskai \cite{KingR04} proved the above conjecture for the special
case of $p=2$ when one of the super-operators satisfies a certain technical
condition that implies multiplicativity for some interesting examples of
channels.

\subsection*{Statement of results}

Although the above super-operator norms have been considered previously
in the context of quantum information theory, many very basic questions about
them have apparently not been addressed.
The purpose of this paper is to investigate some of these basic questions,
and to prove various properties of the norms $\|\Phi\|_{q\rightarrow p}$ and
$\|\Phi\|_{q\rightarrow p}^H$.
The following facts are proved.

\begin{mylist}{\parindent}
\item[1.]
If $\Phi$ is completely positive, the value of the supremum in the definition
of $\|\Phi\|_{q\rightarrow p}$ is achieved by a positive semidefinite
operator $X$, and therefore
$\|\Phi\|_{q\rightarrow p} = \|\Phi\|_{q\rightarrow p}^H$.
This holds for all choices of $p,q\in [1,\infty]$, answering a question posed
by King and Ruskai~\cite{KingR04}.

\item[2.]
For any choice of $p\in [1,\infty]$, there exists a super-operator $\Phi$ that
is the difference of two completely positive, trace-preserving super-operators
such that all Hermitian $X$ fail to achieve the supremum in the definition of
$\|\Phi\|_{1\rightarrow p}$.
Therefore, item 1 does not extend to differences of completely positive
super-operators even for the case $q=1$.

\item[3.]
For all $p\geq 2$, $q\leq 2$, and arbitrary $\Phi$, 
the norm $\|\Phi \|_{q\rightarrow p}$ is stable under tensoring $\Phi$ with
the identity super-operator, i.e.,
$\|\Phi \|_{q\rightarrow p} = \|\Phi \otimes I\|_{q\rightarrow p}$.

\item[4.]
For $1\leq p < 2$, the norm $\|\Phi\|_{1\rightarrow p}$ may fail to be stable
with respect to tensoring $\Phi$ with the identity super-operator as described
in item 3, but $\|\Phi\otimes I\|_{1\rightarrow p}$ is stable in this sense
for $I$ the identity super-operator on $\mathrm{L}(\mathcal{H})$
for $\op{dim}(\mathcal{H}) = \op{dim}(\mathcal{F})$.
This generalizes and simplifies a proof of Kitaev \cite{Kitaev97} that
established this fact for the case $p=1$.
(Independently, Gilchrist, Langford, and Nielsen \cite{GilchristL+04}
have given a proof of a closely related fact based on a similar principle.)
\end{mylist}

\noindent
Included in the conclusion section of the paper are a few questions
that may represent helpful challenge problems for readers interested in 
learning more about these super-operator norms.


\section{Preliminaries}
\label{sec:preliminaries}

In this section we will briefly discuss some of the notation used in this
paper and review some previously known facts that will be used.
The books of Bhatia~\cite{Bhatia97} and Kitaev, Shen, and
Vyalyi~\cite{KitaevS+02} may be helpful as general references for
the topics discussed in this paper, and may be consulted for proofs of
facts not explicitly referenced.

All vector spaces considered will be assumed to be finite-dimensional vector
spaces over the complex numbers, and this assumption will not be made explicit
hereafter.
(This is the case of primary interest in quantum information theory.)
Vector spaces will be denoted by scripted letters such as $\mathcal{F}$,
$\mathcal{G}$, etc.
As already mentioned in the introduction, for given $\mathcal{F}$ and
$\mathcal{G}$ the spaces $\mathrm{L}(\mathcal{F},\mathcal{G})$,
$\mathrm{L}(\mathcal{F})$, and $\mathrm{T}(\mathcal{F},\mathcal{G})$ are the
spaces of all linear maps from $\mathcal{F}$ to $\mathcal{G}$, from
$\mathcal{F}$ to itself, and from $\mathrm{L}(\mathcal{F})$ to
$\mathrm{L}(\mathcal{G})$, respectively.
Elements of $\mathrm{T}(\mathcal{F},\mathcal{G})$ will be referred to as
super-operators.

A super-operator $\Phi\in\mathrm{T}(\mathcal{F},\mathcal{G})$ is
positive if $\Phi(X)$ is positive semidefinite whenever $X$ is
positive semidefinite, and is completely positive if
$\Phi\otimes I_{\mathrm{L}(\mathcal{H})} \in 
\mathrm{T}(\mathcal{F}\otimes\mathcal{H},\mathcal{G}\otimes\mathcal{H})$
is positive for all choices of the space $\mathcal{H}$.
Here, $I_{\mathrm{L}(\mathcal{H})}$ denotes the identity super-operator
mapping $\mathrm{L}(\mathcal{H})$ to itself.
(The identity operator from $\mathcal{F}$ to itself is denoted
$I_{\mathcal{F}}$.)
It is known that $\Phi\in\mathrm{T}(\mathcal{F},\mathcal{G})$ is
completely positive if and only if there exist mappings
$A_1,\ldots,A_N\in\mathrm{L}(\mathcal{F},\mathcal{G})$ for some
$N\in\mathbb{N}$ such that $\Phi(X) = \sum_{i=1}^N A_i X A_i^{\ast}$ for all
$X\in\mathrm{L}(\mathcal{F})$.
An arbitrary super-operator $\Phi\in\mathrm{T}(\mathcal{F},\mathcal{G})$
can be expressed as
$\Phi(X) = \sum_{i=1}^N A_i X B_i^{\ast}$ for all $X\in\mathrm{L}(\mathcal{F})$
for some choice of 
$A_1,\ldots,A_N,B_1,\ldots,B_N\in\mathrm{L}(\mathcal{F},\mathcal{G})$
for some $N\in\mathbb{N}$.
In both cases one may take $N = \op{dim}(\mathcal{F})\cdot\op{dim}(
\mathcal{G})$ without loss of generality.

Recall that the singular value decomposition implies that for any
$A\in\mathrm{L}(\mathcal{F},\mathcal{G})$ of rank $r$, it is possible
to construct orthonormal sets
$\{\ket{u_1},\ldots,\ket{u_r}\}\subset\mathcal{F}$ and
$\{\ket{v_1},\ldots,\ket{v_r}\}\subset\mathcal{G}$, and positive
real numbers $s_1,\ldots,s_r$, that satisfy
\[
A = \sum_{i=1}^r s_i \ket{v_i}\bra{u_i}.
\]
It is typical to order the singular values in decreasing order:
$s_1\geq s_2 \geq \cdots \geq s_r > 0$.
The sum may also be taken over a larger range and zero included as a singular
value when it is notationally convenient.
The singular value decomposition may be applied to a bipartite quantum state
in a similar way, although it is typically called the Schmidt decomposition in
this context---for any
$\ket{\psi}\in\mathcal{G}\otimes\mathcal{F}$, one may write
\[
\ket{\psi} = \sum_{i=1}^r s_i \ket{v_i}\ket{u_i}
\]
for 
$\{\ket{u_1},\ldots,\ket{u_r}\}\subset\mathcal{F}$,
$\{\ket{v_1},\ldots,\ket{v_r}\}\subset\mathcal{G}$, and
$s_1,\ldots,s_r$ precisely as
above.

The Schatten $p$-norm of $A$ satisfies
$\|A\|_p = (s_1^p + \cdots + s_r^p)^{1/p}$ for $1\leq p < \infty$ and
$\|A\|_\infty = s_1$.
It follows that $\|A\|_p \leq \|A\|_q$ whenever $p\geq q$.
The norms $\|\cdot \|_1$, $\|\cdot \|_2$, and $\|\cdot \|_{\infty}$
are commonly called the trace norm, the Frobenius norm, and the
operator norm, and are alternately denoted
$\|\cdot\|_{\mathrm{tr}}$, $\|\cdot\|_{F}$, and $\|\cdot\|$.
In this paper, however, we will simply write $\|\cdot\|_p$ for
whatever value of $p$ is appropriate.

For a given $p\in[1,\infty]$, let $p^{\ast}\in[1,\infty]$ be defined by
the equation
\[
\frac{1}{p} + \frac{1}{p^{\ast}} = 1.
\]
It is the case that
\begin{equation}
\label{eq:dual-norms}
\|X\|_p = \sup\{|\langle Y, X\rangle |\,:\,
Y\in\mathrm{L}(\mathcal{F}),\,\|Y\|_{p^{\ast}}= 1\}
\end{equation}
for every $X\in\mathrm{L}(\mathcal{F})$.
A related fact is that if $X,Y\in\mathrm{L}(\mathcal{F})$ are 
any two operators and $p\in[1,\infty]$, then
\[
|\langle X, Y\rangle| \leq \|X\|_p\, \|Y\|_{p^{\ast}}.
\]

Suppose
\[
X = \sum_{i,j}X_{i,j}\otimes \ket{i}\bra{j}.
\]
Up to a permutation of rows and columns, $X$ may be viewed as a block matrix
with blocks $X_{i,j}$.
Then for $p\in[1,2]$ we have
\[
\sum_{i,j} \|X_{i,j}\|_p^2 \leq \|X\|_p^2
\]
and for $p\in[2,\infty]$ we have
\[
\|X\|_p^2 \leq \sum_{i,j} \|X_{i,j}\|_p^2.
\]
This fact was proved by Bhatia and Kattaneh~\cite{BhatiaK90}.

The norms $\|\Phi\|_{q\rightarrow p}$,
$\|\Phi\|_p=\|\Phi\|_{1\rightarrow p}$,
$\|\Phi\|_{q\rightarrow p}^H$, and
$\|\Phi\|_p^H = \|\Phi\|_{1\rightarrow p}^H$ of
$\Phi\in\mathrm{T}(\mathcal{F},\mathcal{G})$, for any choice of
$p,q\in[1,\infty]$, have been defined in the introduction.
The singular value decomposition along with convexity of norms implies that
for an arbitrary super-operator $\Phi\in\mathrm{T}(\mathcal{F},\mathcal{G})$,
the value $\|\Phi\|_p$ is achieved by
$\|\Phi(\ket{u}\bra{v})\|_p$ for some choice of unit vectors
$\ket{u},\ket{v}\in\mathcal{F}$.
Using the spectral decomposition in place of the singular value
decomposition, we have
\[
\|\Phi\|_p^H = \|\Phi(\ket{u}\bra{u})\|_p^H
\]
for some unit vector $\ket{u}\in\mathcal{F}$.

For spaces $\mathcal{F}$, $\mathcal{G}$, and $\mathcal{H}$, and any completely
positive super-operator $\Phi\in\mathrm{T}(\mathcal{F},\mathcal{G})$, it holds 
that
\[
\|\Phi\|_p^H = \|\Phi\otimes I_{\mathrm{L}(\mathcal{H})}\|^H_p
\]
for all $p\in[1,\infty]$.
This fact was proved by Amosov, Holevo, and Werner \cite{AmosovH+00}.


\section{Hermitian super-operator norms}

This section focuses on the difference between 
$\|\Phi\|_{q\rightarrow p}$ and $\|\Phi\|_{q\rightarrow p}^H$ for
different classes of super-operators $\Phi$.
It is obvious that without any restrictions on $\Phi$, the
quantities $\|\Phi\|_{q\rightarrow p}$ and $\|\Phi\|_{q\rightarrow p}^H$
can differ significantly.
For instance, let $\mathcal{F}$ and $\mathcal{G}$ be two-dimensional spaces
both having standard basis $\{\ket{0},\ket{1}\}$, and let
\[
\Phi(X) = \ket{0}\bra{0}X\ket{1}\bra{0}.
\]
Then $\|\Phi\|_{q\rightarrow p} = 1$ (for all choices of $p,q\in [1,\infty]$),
while $\|\Phi\|_{q\rightarrow p}^H = 2^{-1/q} < 1$ for $q<\infty$.
For $q = \infty$, consider
\[
\Phi(X) = \frac{1}{2}\ket{0}\bra{0}X\ket{0}\bra{0} 
+ \frac{i}{2} \ket{0}\bra{1}X\ket{1}\bra{0}.
\]
Then $\|\Phi\|_{\infty\rightarrow p} = 1$ while
$\|\Phi\|_{\infty\rightarrow p}^H = 1/\sqrt{2}$ (for all $p\in[1,\infty]$).

King and Ruskai \cite{KingR04} raised the question of whether a strict
inequality $\|\Phi\|_{q\rightarrow p}^H<\|\Phi\|_{q\rightarrow p}$ may
occur for some choice of $p,q\in [1,\infty]$ when $\Phi$ is completely
positive.
We prove that this is not possible.

\vspace{2mm}

\begin{theorem}
\label{theorem:Hermitian}
Let $\Phi\in\mathrm{T}(\mathcal{F},\mathcal{G})$ be completely positive.
Then for all $p,q\in [1,\infty]$,
\[
\|\Phi\|_{q\rightarrow p} = \|\Phi\|_{q\rightarrow p}^H.
\]
\end{theorem}

\vspace{2mm}

\noindent
The following lemma establishes an inequality from which this theorem
will follow.

\vspace{2mm}

\begin{lemma}
\label{lemma:Hermitian}
Let $A_1,\ldots,A_N,B_1,\ldots,B_N\in\mathrm{L}(\mathcal{F},\mathcal{G})$
be linear mappings and let
$\Phi\in\mathrm{T}(\mathcal{F},\mathcal{G})$ be given by
\[
\Phi(X) = \sum_{i = 1}^N A_i X B_i^{\ast}
\]
for all $X\in\mathrm{L}(\mathcal{F})$.
Define $\Phi_L,\Phi_R\in\mathrm{T}(\mathcal{F},\mathcal{G})$ as
\[
\Phi_L (X) = \sum_{i = 1}^N A_i X A_i^{\ast},\;\;\;\;
\Phi_R (X) = \sum_{i = 1}^N B_i X B_i^{\ast}.
\]
Then
\[
\| \Phi \|_{q\rightarrow p} \leq 
\sqrt{\| \Phi_L \|_{q\rightarrow p}^H}
\sqrt{\| \Phi_R \|_{q\rightarrow p}^H}
\]
for any choice of $p,q\in[1,\infty]$.
\end{lemma}

\vspace{2mm}

\noindent
{\bf Proof.}
Let $X\in\mathrm{L}(\mathcal{F})$ and $Y\in\mathrm{L}(\mathcal{G})$
satisfy $\|X\|_q = \|Y\|_{p^{\ast}} = 1$, and let
\[
X = \sum_{i=1}^n s_i \ket{u_i}\bra{v_i}\;\;\;\;\mbox{and}\;\;\;\;
Y = \sum_{j=1}^m t_j \ket{w_j}\bra{x_j}
\]
be singular value decompositions of $X$ and $Y$.
Let $X_L,X_R\in\mathrm{L}(\mathcal{F})$ and
$Y_L,Y_R\in\mathrm{L}(\mathcal{G})$ be defined as
\[
X_L = \sum_{i = 1}^n s_i \ket{u_i}\bra{u_i},\;\;
X_R = \sum_{i = 1}^n s_i \ket{v_i}\bra{v_i},\;\;
Y_L = \sum_{i = 1}^m t_i \ket{w_i}\bra{w_i},\;\;
Y_R = \sum_{i = 1}^m t_i \ket{x_i}\bra{x_i}.
\]
Equivalently, $X_L = \sqrt{X X^{\ast}}$, $X_R = \sqrt{X^{\ast} X}$,
$Y_L = \sqrt{Y Y^{\ast}}$, and $Y_R = \sqrt{Y^{\ast} Y}$.
Each of these operators is positive semidefinite.
As $X$, $X_L$, and $X_R$ share the same singular values
$s_1,\ldots, s_n$, we have
\[
\|X_L\|_q = \|X_R\|_q = \|X\|_q = 1,
\]
and similarly
\[
\|Y_L\|_{p^{\ast}} = \|Y_R\|_{p^{\ast}} = \|Y\|_{p^{\ast}} = 1.
\]

Now,
\begin{align*}
|\langle Y, \Phi(X)\rangle|
& = \left|\sum_{i=1}^n\sum_{j=1}^m\sum_{k=1}^N
s_i t_j \opbracket{w_j}{A_k}{u_i}\opbracket{v_i}{B_k^{\ast}}{x_j}\right|\\
& \leq
\sqrt{\sum_{i=1}^n\sum_{j=1}^m\sum_{k=1}^{N}
s_i t_j \left|\opbracket{w_j}{A_k}{u_i}\right|^2}
\sqrt{\sum_{i=1}^n\sum_{j=1}^m\sum_{k=1}^{N}
s_i t_j \left|\opbracket{x_j}{B_k}{v_i}\right|^2}\\
& = \sqrt{\langle Y_L,\Phi_L(X_L)\rangle}
\sqrt{\langle Y_R,\Phi_R(X_R)\rangle}\\
& \leq \sqrt{\|\Phi_L\|_{q\rightarrow p}^H}
\sqrt{\|\Phi_R\|_{q\rightarrow p}^H}.
\end{align*}
The first inequality follows from the Cauchy-Schwarz inequality,
and the second follows from Eq.~\ref{eq:dual-norms} along with the
fact that $X_L$ and $X_R$ are Hermitian.
Taking the supremum over all choices of $X$ and $Y$ with
$\|X\|_q = \|Y\|_{p^{\ast}} = 1$ proves the lemma.
\qed

It should be noted that $\Phi_L$ and $\Phi_R$ as defined in the previous
lemma may depend on the choice of $A_1,\ldots,A_N$ and $B_1,\ldots,B_N$
for a given~$\Phi$, and so they are not well-defined given only~$\Phi$.

\vspace{2mm}

\noindent
{\bf Proof of Theorem~\ref{theorem:Hermitian}}.
As $\Phi$ is completely positive, we may write
$\Phi(X) = \sum_{i=1}^N A_i X A_i^{\ast}$ for some choice of
$A_1,\ldots,A_N\in\mathrm{L}(\mathcal{F},\mathcal{G})$.
We then have $\Phi_L = \Phi_R = \Phi$, and so
$\|\Phi\|_{q\rightarrow p} \leq \|\Phi\|_{q\rightarrow p}^H$ by
Lemma~\ref{lemma:Hermitian}.
\qed

The previous theorem suggests the following question: under what conditions
on $\Phi$ does it hold that
$\|\Phi\|_{q\rightarrow p} = \|\Phi\|_{q\rightarrow p}^H$?
For example, if $\Phi$ is the difference of two completely positive
super-operators,
is it necessary that $\|\Phi\|_{q\rightarrow p} = \|\Phi\|_{q\rightarrow p}^H$?
We prove that this is not a sufficient condition.
Our counter-examples are restricted to the case where $q = 1$.

\vspace{2mm}

\begin{prop}
For any choice of $p\in [1,\infty]$ there exist completely positive
trace-preserving super-operators
$\Phi_0,\Phi_1\in\mathrm{T}(\mathcal{F},\mathcal{G})$
such that
\[
\left\| \Phi_0 - \Phi_1\right\|_p^H < \left\| \Phi_0 - \Phi_1\right\|_p.
\]
\end{prop}

\vspace{2mm}

\noindent
{\bf Proof.}
For $1< p\leq \infty$, the proposition is quite straightforward.
Let $\mathcal{F}$ and $\mathcal{G}$ both have dimension 2, let $\Phi_0$ be the
identity super-operator, and let $\Phi_1$ be defined by 
$\Phi_1(X) = \frac{\op{tr}(X)}{2} I$.
For any unit vector $\ket{\psi}\in\mathcal{F}$, 
\[
\Phi(\ket{\psi}\bra{\psi}) = 
\Phi_0(\ket{\psi}\bra{\psi}) - \Phi_1(\ket{\psi}\bra{\psi}) = 
\ket{\psi}\bra{\psi} - \frac{1}{2}I
\]
has two singular values both equal to 1/2, and thus
$\|\Phi\|_p^H = 2^{1/p}/2 < 1$.
However, for orthogonal unit vectors $\ket{\psi}$ and $\ket{\phi}$,
$\Phi(\ket{\psi}\bra{\phi}) = \ket{\psi}\bra{\phi}$, which implies
$\|\Phi\|_p = 1$.

For $p = 1$ our counter-example is slightly more complicated.
Let $\mathcal{F}$ be a space of dimension 2 and let $\mathcal{G}$
be a space of dimension 4.
The standard bases of these spaces will be written $\{\ket{0},\ket{1}\}$ and
$\{\ket{0},\ket{1}, \ket{2},\ket{3}\}$, respectively.
Define unit vectors $\ket{+},\ket{-}\in\mathcal{F}$ as
\begin{align*}
\ket{+} & = \frac{1}{\sqrt{2}}\ket{0} + \frac{1}{\sqrt{2}}\ket{1}\\
\ket{-} & = \frac{1}{\sqrt{2}}\ket{0} - \frac{1}{\sqrt{2}}\ket{1},
\end{align*}
and define $\Phi_0,\Phi_1\in\mathrm{T}(\mathcal{F},\mathcal{G})$ by
\begin{align*}
\Phi_0 (X) & = \frac{1}{2}\left(
\ket{0}\bra{0}X\ket{0}\bra{0}
+ \ket{1}\bra{+}X\ket{+}\bra{1}
+ \ket{2}\bra{1}X\ket{1}\bra{2}
+ \ket{3}\bra{-}X\ket{-}\bra{3}\right),\\
\Phi_1 (X) & = \frac{1}{2}\left(
\ket{0}\bra{1}X\ket{1}\bra{0}
+ \ket{1}\bra{-}X\ket{-}\bra{1}
+ \ket{2}\bra{0}X\ket{0}\bra{2}
+ \ket{3}\bra{+}X\ket{+}\bra{3}\right)
\end{align*}
for each $X\in\mathrm{L}(\mathcal{F})$.
It is evident that $\Phi_0$ and $\Phi_1$ are completely positive and
trace-preserving.
Finally, let $\Phi = \Phi_0 - \Phi_1$.

For any unit vector $\ket{\psi}\in\mathcal{F}$ we have
\begin{align*}
\Phi(\ket{\psi}\bra{\psi}) & =
\frac{1}{2}\left(|\bracket{0}{\psi}|^2 - |\bracket{1}{\psi}|^2\right)
\ket{0}\bra{0} +
\frac{1}{2}\left(|\bracket{+}{\psi}|^2 - |\bracket{-}{\psi}|^2\right)
\ket{1}\bra{1}\\
& +
\frac{1}{2}\left(|\bracket{1}{\psi}|^2 - |\bracket{0}{\psi}|^2\right)
\ket{2}\bra{2}
+
\frac{1}{2}\left(|\bracket{-}{\psi}|^2 - |\bracket{-}{\psi}|^2\right)
\ket{3}\bra{3},
\end{align*}
and thus
\[
\|\Phi(\ket{\psi}\bra{\psi})\|_1 =
\left| |\bracket{0}{\psi}|^2 - |\bracket{1}{\psi}|^2 \right|
+
\left| |\bracket{+}{\psi}|^2 - |\bracket{-}{\psi}|^2 \right|.
\]
No choice of a unit vector $\ket{\psi}\in\mathcal{F}$ can simultaneously
satisfy both
$\left| |\bracket{0}{\psi}|^2 - |\bracket{1}{\psi}|^2 \right| = 1$
and 
$\left| |\bracket{+}{\psi}|^2 - |\bracket{-}{\psi}|^2 \right|=1$,
which implies $\|\Phi(\ket{\psi}\bra{\psi})\|_1 < 2$.
However, if we define
\begin{align*}
\ket{\circlearrowright} & = \frac{1}{\sqrt{2}}\ket{0} +
\frac{i}{\sqrt{2}}\ket{1}\\
\ket{\circlearrowleft} & = \frac{1}{\sqrt{2}}\ket{0} -
\frac{i}{\sqrt{2}}\ket{1}
\end{align*}
and consider
\[
\Phi(\ket{\circlearrowright}\bra{\circlearrowleft})
= \frac{1}{2}\ket{0}\bra{0}
+ \frac{i}{2}\ket{1}\bra{1}
- \frac{1}{2}\ket{2}\bra{2}
- \frac{i}{2}\ket{3}\bra{3},
\]
then we see that 
$\|\Phi(\ket{\circlearrowright}\bra{\circlearrowleft})\|_1 = 2$.
This implies $\|\Phi\|_1^H < \|\Phi\|_1$ as claimed.
\qed


\section{Stabilizations of super-operator norms}

In this section we consider norms of super-operators tensored with the
identity super-operator.
Suppose $\Phi\in\mathrm{T}(\mathcal{F},\mathcal{G})$ is an arbitrary
super-operator, and $\mathcal{H}$ is a vector space with dimension at least 2.
For $p\in [1,2)$ it may happen that
$\|\Phi\|_p < \|\Phi\otimes I_{\mathrm{L}(\mathcal{H})}\|_p$.
In particular, for $\mathcal{F}$ and $\mathcal{H}$ spaces of dimension $n$
and $T\in\mathrm{L}(\mathcal{F})$ representing matrix transposition with
respect to any orthonormal basis of $\mathcal{F}$, we have $\|T\|_p = 1$
while
\[
\|T\otimes I_{\mathrm{L}(\mathcal{H})}\|_p = \frac{n^{2/p}}{n} > 1.
\]
For $p\geq 2$, however, this phenomenon does not occur.
More generally, this statement holds for any choice of $q\leq 2$ in place of
$q = 1$.

\vspace{2mm}

\noindent
\begin{theorem}
\label{theorem:stable1}
Let $\mathcal{F}$, $\mathcal{G}$, and $\mathcal{H}$ be finite dimensional
spaces and let $\Phi\in\mathrm{T}(\mathcal{F},\mathcal{G})$ be an arbitrary
super-operator.
Then for $p\geq 2$ and $q\leq 2$,
\[
\|\Phi\|_{q\rightarrow p} = 
\|\Phi\otimes I_{\mathrm{L}(\mathcal{H})}\|_{q\rightarrow p}.
\]
\end{theorem}

\vspace{2mm}

\noindent
{\bf Proof.}
It suffices to prove that
$\|\Phi\|_{q\rightarrow p} \geq 
\|\Phi\otimes I_{\mathrm{L}(\mathcal{H})}\|_{q\rightarrow p}$,
as the reverse inequality is straightforward.
Recall that $p^{\ast}$ is defined by the equation $1/p + 1/p^{\ast} = 1$,
which implies that $p^{\ast} \leq 2$.

Let $X\in\mathrm{L}(\mathcal{F}\otimes\mathcal{H})$
and $Y\in\mathrm{L}(\mathcal{G}\otimes\mathcal{H})$
satisfy $\|X\|_q = \|Y\|_{p^{\ast}} = 1$.
Write
\[
X = \sum_{i,j = 1}^k X_{i,j}\otimes \ket{i}\bra{j}\;\;\;\mbox{and}\;\;\;
Y = \sum_{i,j = 1}^k Y_{i,j}\otimes \ket{i}\bra{j}
\]
for $k = \op{dim}(\mathcal{H})$ and $X_{i,j} \in\mathrm{L}(\mathcal{G})$,
$Y_{i,j} \in\mathrm{L}(\mathcal{F})$ for $1\leq i,j\leq k$.
As $p^{\ast},q\in[1,2]$ we have
\[
\sum_{i,j} \|X_{i,j}\|_q^2 \leq \|X\|_q^2 = 1\;\;\;
\mbox{and}\;\;\;
\sum_{i,j} \|Y_{i,j}\|_{p^{\ast}}^2 \leq \|Y\|_{p^{\ast}}^2 = 1
\]
as noted in Section~\ref{sec:preliminaries}.
Now,
\begin{align*}
\left|\langle Y, (\Phi\otimes I_{\mathrm{L}(\mathcal{H})})(X) \rangle \right|
& =
\left|\sum_{i,j}\left\langle Y_{i,j}, \Phi(X_{i,j})\right\rangle\right|\\
& \leq \sum_{i,j} \|Y_{i,j}\|_{p^{\ast}} \|\Phi(X_{i,j})\|_p\\
& \leq \|\Phi\|_{q\rightarrow p}\sum_{i,j}\|Y_{i,j}\|_{p^{\ast}}\|X_{i,j}\|_q\\
& \leq \|\Phi\|_{q\rightarrow p} \sqrt{\sum_{i,j} \|Y_{i,j}\|_{p^{\ast}}^2}
\sqrt{\sum_{i,j} \|X_{i,j}\|_q^2}\\
& \leq \|\Phi\|_{q\rightarrow p}.
\end{align*}
Taking the supremum over all $Y$ with $\|Y\|_{p^{\ast}} = 1$
establishes that $\|(\Phi\otimes I_{\mathrm{L}(\mathcal{H})})(X)\|_p \leq
\|\Phi\|_{q\rightarrow p}$ for all $X$ with $\|X\|_q = 1$, and thus 
$\|\Phi\otimes I_{\mathrm{L}(\mathcal{H})}\|_{q\rightarrow p}
\leq \|\Phi\|_{q\rightarrow p}$ as required.
\qed

Although the previous theorem is not true in the case $p<2$ and $q = 1$,
there is a limit to the possible increase as the dimension of
$\mathcal{H}$ increases.
In particular, the increase cannot continue after the dimension
of $\mathcal{H}$ reaches that of the input space $\mathcal{F}$.
Kitaev \cite{Kitaev97} proved this fact for the case $p=1$.
The next theorem generalizes this fact to all $p$.
(Of course this fact follows trivially from Theorem~\ref{theorem:stable1}
for $p\geq 2$, but the proof works for arbitrary $p$.)

\vspace{2mm}

\noindent
\begin{theorem}
Let $\mathcal{F}$, $\mathcal{G}$, $\mathcal{H}$, and $\mathcal{K}$
be finite-dimensional spaces with $\op{dim}(\mathcal{H})\geq
\op{dim}(\mathcal{K}) = \op{dim}(\mathcal{F})$, and let
$\Phi\in\mathrm{T}(\mathcal{F},\mathcal{G})$ be an
arbitrary super-operator.
Then for all $p\in[1,\infty]$,
\[
\|\Phi\otimes I_{\mathrm{L}(\mathcal{H})}\|_p =
\|\Phi\otimes I_{\mathrm{L}(\mathcal{K})}\|_p\;\;\;\mbox{and}\;\;\;
\|\Phi\otimes I_{\mathrm{L}(\mathcal{H})}\|_p^H =
\|\Phi\otimes I_{\mathrm{L}(\mathcal{K})}\|_p^H.
\]
\end{theorem}

\vspace{2mm}
\noindent
{\bf Proof.}
We have that
\[
\|\Phi \otimes I_{\mathrm{L}(\mathcal{H})}\|_p
=
\|(\Phi \otimes I_{\mathrm{L}(\mathcal{H})})(\ket{u}\bra{v})\|_p
\]
for some choice of unit vectors
$\ket{u},\ket{v}\in\mathcal{F}\otimes\mathcal{H}$.
Fix such a choice of $\ket{u}$ and $\ket{v}$, and let
\[
\ket{u} = \sum_{i=1}^n s_i \ket{w_i}\ket{y_i}\;\;\;\mbox{and}\;\;\;
\ket{v} = \sum_{i=1}^n t_i \ket{x_i}\ket{z_i}
\]
be Schmidt decompositions of $\ket{u}$ and $\ket{v}$, where
$n = \op{dim}(\mathcal{F}) = \op{dim}(\mathcal{K})$.
Then
\[
(\Phi\otimes I_{\mathrm{L}(\mathcal{H})})(\ket{u}\bra{v})
=\sum_{i,j}s_i t_j \Phi(\ket{w_i}\bra{x_j}) \otimes \ket{y_i}\bra{z_j}.
\]
Let $\{\ket{i}\,:\,i=1,\ldots,n\}$ represent an orthonormal basis of
$\mathcal{K}$ and define $U,V\in\mathrm{L}(\mathcal{H},\mathcal{K})$ as
\[
U = \sum_{i=1}^n \ket{i}\bra{y_i}\;\;\;\mbox{and}\;\;\;
V = \sum_{i=1}^n \ket{i}\bra{z_i}.
\]
The mappings $U^{\ast}U$ and $V^{\ast}V$ are projections onto
the spaces spanned by $\ket{y_1},\ldots,\ket{y_n}$ and
$\ket{z_1},\ldots,\ket{z_n}$, respectively.
It therefore follows that
$\|U\|_{\infty} = \|V\|_{\infty} = 1$ and that 
$I\otimes U^{\ast}U$ and $I\otimes V^{\ast}V$ act trivially on
$\ket{u}$ and $\ket{v}$, respectively.
We then have
\begin{align*}
\|\Phi\otimes I_{\mathrm{L}(\mathcal{K})}\|_p & \geq
\left\|
(\Phi\otimes I_{\mathrm{L}(\mathcal{K})})
((I\otimes U)\ket{u}\bra{v}(I\otimes V^{\ast}))
\right\|_p\\
& \geq
\left\|(I\otimes U^{\ast})
(\Phi\otimes I_{\mathrm{L}(\mathcal{K})})
((I\otimes U)\ket{u}\bra{v}(I\otimes V^{\ast}))
(I\otimes V)
\right\|_p\\
& = 
\left\|
(\Phi\otimes I_{\mathrm{L}(\mathcal{H})})
((I\otimes U^{\ast}U)\ket{u}\bra{v}(I\otimes V^{\ast}V))
\right\|_p\\
& = 
\left\|
(\Phi\otimes I_{\mathrm{L}(\mathcal{H})})
(\ket{u}\bra{v})\right\|_p\\
& = \|\Phi\otimes I_{\mathrm{L}(\mathcal{H})}\|_p.
\end{align*}

To prove 
$\|\Phi\otimes I_{\mathrm{L}(\mathcal{H})}\|_p^H =
\|\Phi\otimes I_{\mathrm{L}(\mathcal{K})}\|_p^H$,
the same argument applies, using the additional assumption
$\ket{u} = \ket{v}$ (and therefore $U = V$).
\qed


\section{Conclusion}

The purpose of this paper has been to investigate super-operator norms induced
by Schatten norms, and to establish some basic properties of these norms.
Possible applications of these facts have not been considered in this paper,
but the study of these norms is justifiable given their connections to
fundamental open problems in quantum information theory.
We conclude with some questions about these norms that may be helpful
stimulating further research on this topic.
\begin{mylist}{\parindent}
\item[1.]
For an arbitrary super-operator $\Phi\in\mathrm{T}(\mathcal{F},\mathcal{G})$
and an arbitrary space $\mathcal{H}$, we have
\[
\|\Phi\otimes I_{\mathrm{L}(\mathcal{H})}\|_p^2
\leq \|\Phi_L\|_p \,\|\Phi_R\|_p,
\]
where $\Phi_L$ and $\Phi_R$ satisfy the conditions of
Lemma~\ref{lemma:Hermitian}.
For $p = 1$ and $\op{dim}(\mathcal{H})\geq\op{dim}(\mathcal{F})$ it in
fact holds that
\[
\|\Phi\otimes I_{\mathrm{L}(\mathcal{H})}\|_1^2
= \inf\{\|\Phi_L\|_1 \|\Phi_R\|_1\},
\]
where the infimum is over all possible $\Phi_L$ and $\Phi_R$ satisfying the
required conditions.
(This follows by an alternate characterization of the $\|\cdot\|_{\Diamond}$
norm proved by Kitaev~\cite{Kitaev97}.)
Does this fact hold for any (or all) values of $p>1$?

\item[2.]
We have proved that if $\Phi\in\mathrm{T}(\mathcal{F},\mathcal{G})$ and
$\op{dim}(\mathcal{H}) \geq \op{dim}(\mathcal{F})$, then
\[
\|\Phi\otimes I_{\mathrm{L}(\mathcal{H})}\|_p = 
\|\Phi\otimes I_{\mathrm{L}(\mathcal{F})}\|_p.
\]
Is it the case that
\[
\|\Phi\otimes I_{\mathrm{L}(\mathcal{H})}\|_{q\rightarrow p} = 
\|\Phi\otimes I_{\mathrm{L}(\mathcal{F})}\|_{q\rightarrow p}
\]
for $1\leq q \leq p$?
This equality is not true in general for $q>p$.

\item[3.]
If $\Phi\in\mathrm{T}(\mathcal{F},\mathcal{G})$ is completely positive,
then $\|\Phi\|_p = \|\Phi\otimes I\|_p$ for $I$ the identity super-operator
on an arbitrary space.
Does this hold for
$\|\Phi\|_{q\rightarrow p}$ versus $\|\Phi\otimes I\|_{q\rightarrow p}$
for $1\leq q\leq p$?
For $q\leq 2$ and $p\geq 2$ we have shown that this holds (even without
the completely positive condition), while again equality is easily seen
not to be true for $q > p$.
\end{mylist}


\subsection*{Acknowledgments}

This research was supported by Canada's NSERC,
the Canadian Institute for Advanced Research (CIAR),
and the Canada Research Chairs Program.


\bibliographystyle{plain}


\end{document}